\documentclass[10pt,a4paper,twoside]{article}
\usepackage{epsfig}
\usepackage{baltlat6}
\usepackage{array}
\usepackage{here}
\begin{document}
\ \
\vspace{0.5mm}
\setcounter{page}{580}
\vspace{8mm}

\titlehead{Baltic Astronomy, vol.\,20, 580--586, 2011}

\titleb{STARK BROADENING OF SEVERAL Ne II, Ne III  AND O III 
\\ SPECTRAL LINES FOR THE STARK-B DATABASE}

\begin{authorl}
\authorb{Milan S. Dimitrijevi\'{c}}{1,2},
\authorb{Andjelka Kova\v cevi\'{c}}{3},
\authorb{Zoran Simi\'{c}}{1} \\ and
\authorb{Sylvie Sahal-Br\'{e}chot}{2}
\end{authorl}

\begin{addressl}
\addressb{1}{Astronomical Observatory, Volgina 7, 11060
Belgrade 38, Serbia;\\
mdimitrijevic@aob.bg.ac.rs, zsimic@aob.bg.ac.rs}
\addressb{2}{Observatoire de Paris, LERMA, 5 Place Jules Janssen, 92190 Meudon, France; sylvie.sahal-brechot@obspm.fr}
\addressb{3}{Department of Astronomy, Faculty of Mathematics, Studentski Trg 15, 11000  Belgrade, Serbia; andjelka@matf.bg.ac.rs}
\end{addressl}

\submitb{Received: 2011 August 8; accepted: 2011 August 15}

\begin{summary} In order to complete Stark broadening data for Ne II, and O III lines, needed for analysis of stellar atmospheres, we determined, within the semiclassical perturbation method, the missing Stark broadening parameters for the broadening by collisions with protons and ionized helium, for 15 Ne II and 5 O III multiplets. Also, electron, proton, and ionized
helium impact broadening parameters for an important Ne II multiplet in the visible part of the spectrum, and for three Ne III multiplets, were calculated. The obtained data will be included in the   STARK-B database, which is a part of Virtual Atomic and Molecular Data Center.
\end{summary}


\resthead {Stark broadening of Ne II, Ne III  and O III spectral lines}
{M. S. Dimitrijevi\'{c}, A. Kova\v {c}evi\'{c}, Z. Simi\'{c}, S. Sahal-Br\'echot}

\sectionb{1}{INTRODUCTION}
The data on Stark broadening of spectral lines, are of interest for
diagnostics, modelling and investigations of stellar atmospheres and other various plasmas in
astrophysics, laboratory, technology and fusion research.
Such data obtained by us using a semiclassical
perturbation method, are organized in the STARK-B database (http://stark-b.obspm.fr/), a part of Virtual Atomic and Molecular Data Center (VAMDC - http://vamdc.org/, Dubernet et al., 2010, Rixon et al., 2011),
supported by EU in the framework of the FP7 Research Infrastructures-INFRA-2008-1.2.2-Scientific Data Infrastructures initiative.

\begin{table}[!t]
\begin{center}
\vbox{\footnotesize\tabcolsep=3pt
\parbox[c]{124mm}{\baselineskip=10pt
{\smallbf\ \ Table 1.}{\small\
 Stark full widths at half intensity maximum (W) and shifts (d) due to electron, proton, and ionized helium impacts for Ne II, for a perturber density of 10$^{17}$ cm$^{-3}$. The quantity C (given in $\AA $cm$^{-3}$), when divided by the corresponding perturber width, gives an estimate for the maximum perturber density for which tabulated data may be used.  \lstrut}}
\resizebox{12cm}{!} {
\begin{tabular}{cccccccc}
\hline
      &&   Electrons & & Protons  & &           Ionized helium \\
 TRANSITION &   T(K) &  W(\AA) & d(\AA) & W(\AA) & d(\AA) &  W(\AA) & d(\AA) \hstrut\lstrut\\
\hline

Ne II       &  5000.& 0.376&   0.0257& 0.0151& 0.273E-02& 0.0188& 0.256E-02 \hstrut\\
  3292.1 \AA   & 10000.& 0.281&   0.0149& 0.0224& 0.502E-02& 0.0259& 0.437E-02 \\
 C= 0.11E+21 & 20000.& 0.214&   0.0133& 0.0277& 0.741E-02& 0.0297& 0.641E-02 \\
 3p$^2$D$^{\degr}$-3d$^2$P   & 30000.& 0.190&   0.0141& 0.0300& 0.869E-02& 0.0318& 0.716E-02 \\
             & 50000.& 0.171&   0.0147& 0.0328& 0.100E-01& 0.0341& 0.836E-02 \\
             &100000.& 0.157&   0.0127& 0.0356& 0.120E-01& 0.0357& 0.998E-02\lstrut \\
\hline
\end{tabular}
}}
\end{center}
\vskip-1mm
\end{table}

\begin{table}[]
\begin{center}
\vbox{\footnotesize\tabcolsep=3pt
\parbox[c]{124mm}{\baselineskip=10pt
{\smallbf\ \ Table 2.}{\small\
Stark broadening parameters for Ne II multiplets for a perturber
 density of 10$^{17}$ cm$^{-3}$. Designations are the same as in Table 1.\lstrut}}
\resizebox{9cm}{!} {
\begin{tabular}{cccccc}
\hline
      & &   Protons  & &           Ionized &helium \\
 TRANSITION &   T(K) &  W(\AA) & d(\AA) &  W(\AA) & d(\AA) \hstrut\lstrut\\
\hline

  Ne II      &   5000.&  0.152E-01&  0.276E-02&  0.190E-01&  0.260E-02 \hstrut\\
  3375.6 \AA   &  10000.&  0.227E-01&  0.510E-02&  0.264E-01&  0.445E-02 \\
 C= 0.13E+21 &  20000.&  0.282E-01&  0.755E-02&  0.303E-01&  0.653E-02 \\
 3p$^2$D$^{\degr}$-3d$^2$F   &  30000.&  0.305E-01&  0.888E-02&  0.324E-01&  0.729E-02 \\
             &  50000.&  0.333E-01&  0.102E-01&  0.348E-01&  0.853E-02 \\
             & 100000.&  0.363E-01&  0.122E-01&  0.365E-01&  0.101E-01 \\
\hline

 Ne II       &   5000.&  0.498E-02& -0.512E-03&  0.716E-02& -0.508E-03 \\
  3572.1 \AA   &  10000.&  0.886E-02& -0.109E-02&  0.109E-01& -0.104E-02 \\
 C= 0.36E+21 &  20000.&  0.126E-01& -0.193E-02&  0.140E-01& -0.172E-02 \\
 3s$'^2$D-3p$'^2$F$^{\degr}$ &  30000.&  0.141E-01& -0.236E-02&  0.152E-01& -0.209E-02 \\
             &  50000.&  0.154E-01& -0.304E-02&  0.165E-01& -0.257E-02 \\
             & 100000.&  0.173E-01& -0.371E-02&  0.179E-01& -0.308E-02 \\
\hline

 Ne II       &   5000.&  0.475E-02& -0.200E-03&  0.678E-02& -0.199E-03 \\
  3337.8 \AA   &  10000.&  0.832E-02& -0.439E-03&  0.102E-01& -0.428E-03 \\
 C= 0.33E+21 &  20000.&  0.117E-01& -0.824E-03&  0.130E-01& -0.765E-03 \\
 3s$'^2$D-3p$'^2$P$^{\degr}$ &  30000.&  0.130E-01& -0.109E-02&  0.140E-01& -0.958E-03 \\
             &  50000.&  0.142E-01& -0.141E-02&  0.152E-01& -0.123E-02 \\
             & 100000.&  0.159E-01& -0.182E-02&  0.165E-01& -0.152E-02 \\
\hline

 Ne II       &   5000.&  0.146E-01&  0.221E-02&  0.184E-01&  0.211E-02 \\
  3422.7 \AA   &  10000.&  0.219E-01&  0.417E-02&  0.257E-01&  0.374E-02 \\
 C= 0.14E+21 &  20000.&  0.274E-01&  0.632E-02&  0.295E-01&  0.549E-02 \\
 3p$'^2$P-3d$'^2$P$^{\degr}$ &  30000.&  0.297E-01&  0.754E-02&  0.316E-01&  0.623E-02 \\
             &  50000.&  0.324E-01&  0.876E-02&  0.340E-01&  0.726E-02 \\
             & 100000.&  0.353E-01&  0.104E-01&  0.357E-01&  0.867E-02 \\
\hline

\end{tabular}
}}
\end{center}
\vskip-1mm
\end{table}

\begin{table}[]
\begin{center}
\vbox{\footnotesize\tabcolsep=3pt
\parbox[c]{124mm}{\baselineskip=10pt
{\smallbf\ \ Table 2 continued.}{\small\
 \lstrut}}
\resizebox{9cm}{!} {

\begin{tabular}{cccccc}
\hline
      & &   Protons  & &           Ionized & helium \\
 TRANSITION &   T(K) &  W(\AA) & d(\AA) &  W(\AA) & d(\AA) \hstrut\lstrut\\
\hline

 Ne II       &   5000.&  0.106E-01&  0.195E-02&  0.136E-01&  0.185E-02 \\
  3040.1 \AA   &  10000.&  0.161E-01&  0.364E-02&  0.191E-01&  0.323E-02 \\
 C= 0.11E+21 &  20000.&  0.205E-01&  0.546E-02&  0.220E-01&  0.472E-02 \\
 3p$^4$P$^{\degr}$-3d$^4$D   &  30000.&  0.222E-01&  0.647E-02&  0.236E-01&  0.532E-02 \\
             &  50000.&  0.243E-01&  0.749E-02&  0.254E-01&  0.621E-02 \\
             & 100000.&  0.265E-01&  0.892E-02&  0.268E-01&  0.745E-02 \\
\hline

 Ne II       &   5000.&  0.134E-01&  0.207E-02&  0.170E-01&  0.198E-02 \\
  3352.6 \AA   &  10000.&  0.202E-01&  0.392E-02&  0.238E-01&  0.352E-02 \\
 C= 0.13E+21 &  20000.&  0.255E-01&  0.594E-02&  0.274E-01&  0.517E-02 \\
 3p$^4$D$^{\degr}$-3d$^4$D   &  30000.&  0.276E-01&  0.711E-02&  0.294E-01&  0.587E-02 \\
             &  50000.&  0.301E-01&  0.825E-02&  0.317E-01&  0.685E-02 \\
             & 100000.&  0.329E-01&  0.987E-02&  0.333E-01&  0.815E-02 \lstrut\\
\hline

 Ne II       &   5000.&  0.153E-01&  0.289E-02&  0.192E-01&  0.271E-02 \\
  3421.5 \AA   &  10000.&  0.229E-01&  0.532E-02&  0.267E-01&  0.463E-02 \\
 C= 0.13E+21 &  20000.&  0.285E-01&  0.786E-02&  0.306E-01&  0.680E-02 \\
 3p$^2$D$^{\degr}$-3d$^4$F   &  30000.&  0.309E-01&  0.923E-02&  0.328E-01&  0.759E-02 \\
             &  50000.&  0.338E-01&  0.106E-01&  0.353E-01&  0.888E-02 \\
             & 100000.&  0.368E-01&  0.127E-01&  0.370E-01&  0.105E-01 \lstrut\\
\hline

 Ne II       &   5000.&  0.171E-01&  0.281E-02&  0.212E-01&  0.265E-02 \\
  3473.1 \AA   &  10000.&  0.254E-01&  0.521E-02&  0.292E-01&  0.456E-02 \\
 C= 0.13E+21 &  20000.&  0.312E-01&  0.773E-02&  0.335E-01&  0.668E-02 \\
 3p$^2$S$^{\degr}$-3d$^2$P   &  30000.&  0.337E-01&  0.911E-02&  0.359E-01&  0.749E-02 \\
             &  50000.&  0.368E-01&  0.105E-01&  0.385E-01&  0.872E-02 \\
             & 100000.&  0.398E-01&  0.126E-01&  0.402E-01&  0.103E-01 \lstrut\\
\hline
 
 Ne II       &   5000.&  0.192E-01&  0.281E-02&  0.238E-01&  0.267E-02 \\
  3651.8 \AA   &  10000.&  0.285E-01&  0.526E-02&  0.328E-01&  0.466E-02 \\
 C= 0.14E+21 &  20000.&  0.349E-01&  0.788E-02&  0.375E-01&  0.681E-02 \\
 3p$^2$P$^{\degr}$-3d$^2$P   &  30000.&  0.377E-01&  0.934E-02&  0.401E-01&  0.767E-02 \\
             &  50000.&  0.411E-01&  0.108E-01&  0.430E-01&  0.896E-02 \\
             & 100000.&  0.443E-01&  0.129E-01&  0.450E-01&  0.108E-01 \\
\hline

 Ne II       &   5000.&  0.152E-01&  0.313E-02&  0.190E-01&  0.291E-02 \\
  3352.6 \AA   &  10000.&  0.227E-01&  0.567E-02&  0.263E-01&  0.491E-02 \\
 C= 0.11E+21 &  20000.&  0.282E-01&  0.836E-02&  0.302E-01&  0.717E-02 \\
 3p$^2$D$^{\degr}$-3d$^4$P   &  30000.&  0.306E-01&  0.970E-02&  0.324E-01&  0.801E-02 \\
             &  50000.&  0.334E-01&  0.112E-01&  0.348E-01&  0.930E-02 \\
             & 100000.&  0.362E-01&  0.135E-01&  0.364E-01&  0.110E-01 \\
\hline
\end{tabular}
}}
\end{center}
\vskip-1mm
\end{table}

\begin{table}[]
\begin{center}
\vbox{\footnotesize\tabcolsep=3pt
\parbox[c]{124mm}{\baselineskip=10pt
{\smallbf\ \ Table 2 continued.}{\small\
 \lstrut}}
\resizebox{9cm}{!} {

\begin{tabular}{cccccc}
\hline
      & &   Protons  & &           Ionized & helium \\
 TRANSITION &   T(K) &  W(\AA) & d(\AA) &  W(\AA) & d(\AA) \hstrut\lstrut\\
\hline

 Ne II       &   5000.&  0.131E-01&  0.154E-01&  0.141E-01&  0.129E-01 \\
  3412.3 \AA   &  10000.&  0.229E-01&  0.241E-01&  0.234E-01&  0.202E-01 \\
 C= 0.10E+21 &  20000.&  0.339E-01&  0.323E-01&  0.305E-01&  0.267E-01 \\
 3p$^2$P$^{\degr}$-4s$^2$P   &  30000.&  0.393E-01&  0.362E-01&  0.347E-01&  0.299E-01 \\
             &  50000.&  0.462E-01&  0.418E-01&  0.405E-01&  0.346E-01 \\
             & 100000.&  0.549E-01&  0.486E-01&  0.470E-01&  0.403E-01 \\
\hline

 Ne II       &   5000.&  0.152E-01&  0.190E-02&  0.192E-01&  0.182E-02 \\
  3432.3 \AA   &  10000.&  0.228E-01&  0.362E-02&  0.266E-01&  0.329E-02 \\
 C= 0.14E+21 &  20000.&  0.284E-01&  0.555E-02&  0.305E-01&  0.485E-02 \\
 3p$^2$D$^{\degr}$-3d$^2$D   &  30000.&  0.307E-01&  0.670E-02&  0.327E-01&  0.559E-02 \\
             &  50000.&  0.334E-01&  0.780E-02&  0.351E-01&  0.650E-02 \\
             & 100000.&  0.361E-01&  0.934E-02&  0.366E-01&  0.778E-02 \\
\hline

 Ne II       &   5000.&  0.176E-01&  0.163E-02&  0.221E-01&  0.157E-02 \\
  3650.0 \AA   &  10000.&  0.263E-01&  0.317E-02&  0.306E-01&  0.293E-02 \\
 C= 0.16E+21 &  20000.&  0.325E-01&  0.494E-02&  0.351E-01&  0.437E-02 \\
 3p$^4$S$^{\degr}$-3d$^2$D   &  30000.&  0.351E-01&  0.608E-02&  0.375E-01&  0.517E-02 \\
             &  50000.&  0.382E-01&  0.713E-02&  0.403E-01&  0.596E-02 \\
             & 100000.&  0.412E-01&  0.862E-02&  0.422E-01&  0.716E-02 \\
\hline

   Ne II     &   5000.&  0.177E-01&  0.277E-02&  0.221E-01&  0.263E-02 \\
  3637.8 \AA   &  10000.&  0.264E-01&  0.519E-02&  0.306E-01&  0.460E-02 \\
 C= 0.15E+21 &  20000.&  0.327E-01&  0.778E-02&  0.351E-01&  0.672E-02 \\
 3p$^4$S$^{\degr}$-3d$^4$F   &  30000.&  0.353E-01&  0.921E-02&  0.376E-01&  0.757E-02 \\
             &  50000.&  0.386E-01&  0.107E-01&  0.404E-01&  0.884E-02 \\
             & 100000.&  0.418E-01&  0.127E-01&  0.423E-01&  0.106E-01 \\
\hline

    Ne II    &   5000. & 0.179E-01 & 0.578E-02 & 0.220E-01 & 0.527E-02 \\
  3560.0 \AA   &  10000. & 0.268E-01 & 0.982E-02 & 0.306E-01 & 0.852E-02 \\
 C= 0.12E+21 &  20000.&  0.334E-01&  0.144E-01&  0.353E-01&  0.119E-01 \\
 3p$^4$S$^{\degr}$-3d$^4$P   &  30000.&  0.364E-01&  0.161E-01&  0.379E-01&  0.133E-01 \\
             &  50000.&  0.402E-01&  0.187E-01&  0.411E-01&  0.155E-01 \\
             & 100000.&  0.444E-01&  0.223E-01&  0.435E-01&  0.183E-01\lstrut \\
\hline
\end{tabular}
}}
\end{center}
\vskip-1mm
\end{table}

\begin{table}[]
\begin{center}
\vbox{\footnotesize\tabcolsep=3pt
\parbox[c]{124mm}{\baselineskip=10pt
{\smallbf\ \ Table 3.}{\small\
Stark broadening parameters for Ne III multiplets. for an electron density of 10$^{17}$ cm$^{-3}$. Notation is the same as in Table 1.\lstrut}}
\resizebox{12cm}{!} {

\begin{tabular}{cccccccc}
\hline

      &&   Electrons & & Protons  & &           Ionized &helium \\
 TRANSITION &   T(K) &  W(\AA) & d(\AA) & W(\AA) & d(\AA) &  W(\AA) & d(\AA) \hstrut\lstrut\\
\hline

 Ne III     &  20000.& 0.0931&-0.00108& 0.00182&-0.478E-03& 0.00256&-0.464E-03 \hstrut\\
  2611.8 \AA   & 50000.& 0.0603&-0.00117& 0.00340&-0.102E-02& 0.00412&-0.912E-03 \\
 C= 0.25E+21 &100000.& 0.0458&-0.00157& 0.00451&-0.145E-02& 0.00482&-0.126E-02 \\
3s$'^{3}$D$^{\degr}$-3p$'^3$F  &200000.& 0.0368&-0.00143& 0.00515&-0.183E-02& 0.00542&-0.153E-02 \\
             &300000.& 0.0331&-0.00144& 0.00553&-0.205E-02& 0.00573&-0.170E-02 \\
             &500000.& 0.0294&-0.00136& 0.00597&-0.233E-02& 0.00595&-0.193E-02 \\
\hline

 Ne III     &  20000.& 0.0771&-0.207E-02 &0.00290&-0.477E-04& 0.00376&-0.476E-04 \\
  2265.1 \AA  &50000.& 0.0515&-0.106E-03 &0.00482&-0.120E-03& 0.00544&-0.116E-03 \\
 C= 0.18E+21 &100000.& 0.0398&-0.133E-03 &0.00571&-0.206E-03& 0.00615&-0.189E-03 \\
3p$'^{3}$F-3d$'^{3}$F$^{\degr}$  &200000.& 0.0325&-0.717E-04 &0.00643&-0.299E-03& 0.00681&-0.263E-03 \\
             &300000.& 0.0296&-0.974E-04 &0.00680&-0.361E-03& 0.00702&-0.303E-03 \\
             &500000.& 0.0267&-0.890E-04 &0.00707&-0.416E-03& 0.00723&-0.346E-03 \\
\hline

 Ne III       &20000.& 0.0666&-0.254E-04 &0.00249& 0.306E-04& 0.00323& 0.306E-04 \\
  2092.4 \AA   & 50000.& 0.0443& 0.554E-04 &0.00414& 0.775E-04& 0.00466& 0.753E-04 \\
 C= 0.15E+21& 100000.& 0.0342& 0.833E-04 &0.00490& 0.136E-03& 0.00527& 0.127E-03 \\
 3p$'^{3}$D-3d$'^{3}$D$^{\degr}$& 200000.& 0.0280& 0.120E-03 &0.00551& 0.203E-03& 0.00584& 0.180E-03 \\
            & 300000.& 0.0254& 0.983E-04 &0.00583& 0.248E-03& 0.00601& 0.212E-03 \\
            & 500000.& 0.0230& 0.988E-04 &0.00605& 0.289E-03& 0.00619& 0.242E-03\lstrut \\
\hline
\end{tabular}
}}
\end{center}
\vskip-1mm
\end{table}

\begin{table}[]
\begin{center}
\vbox{\footnotesize\tabcolsep=3pt
\parbox[c]{124mm}{\baselineskip=10pt
{\smallbf\ \ Table 4.}{\small\
Stark broadening parameters for O III multiplets. for an electron density of 10$^{17}$ cm$^{-3}$. Notation is the same as in Table 1. \lstrut}}
\resizebox{9cm}{!} {

\begin{tabular}{cccccc}

\hline
 &&   Protons  & &           Inized &helium \\
 TRANSITION &   T(K) &  W(\AA) & d(\AA) &  W(\AA) & d(\AA) \hstrut\lstrut\\
\hline

  O III     &  20000.&  0.504E-02&-0.213E-02& 0.685E-02&-0.201E-02 \hstrut\\
  3763.4 \AA  &  50000.&  0.921E-02&-0.406E-02& 0.107E-01&-0.356E-02 \\
 C= 0.38E+21& 100000.&  0.118E-01&-0.560E-02& 0.124E-01&-0.469E-02 \\
  3p$^3$D-3s$^3$P$^{\degr}$ & 200000.& 0.137E-01-&0.674E-02  &0.141E-01&-0.564E-02 \\
            & 300000.&  0.148E-01&-0.746E-02 &0.148E-01&-0.624E-02 \\
            & 500000.&  0.161E-01&-0.847E-02 &0.156E-01&-0.700E-02 \\
\hline

  O III     &  20000.& 0.424E-02&-0.148E-02& 0.572E-02&-0.141E-02 \\
  3327.6 \AA  &  50000.& 0.761E-02&-0.289E-02& 0.884E-02&-0.251E-02 \\
 C= 0.29E+21& 100000.& 0.958E-02&-0.397E-02& 0.102E-01&-0.336E-02 \\
  3p$^3$S-3s$^3$P$^{\degr}$ & 200000.& 0.110E-01&-0.483E-02& 0.115E-01&-0.402E-02 \\
            & 300000.& 0.119E-01&-0.537E-02& 0.120E-01&-0.445E-02 \\
            & 500000.& 0.127E-01&-0.612E-02& 0.125E-01&-0.503E-02 \\
\hline

  O III     & 20000. & 0.417E-02&-0.723E-03& 0.552E-02&-0.700E-03 \\
  2984.7 \AA  & 50000. & 0.719E-02&-0.152E-02& 0.823E-02&-0.134E-02 \\
 C= 0.16E+21& 100000.& 0.871E-02&-0.212E-02& 0.934E-02&-0.185E-02 \\
  3p$^1$D-3s$^1$P$^{\degr}$ & 200000.& 0.990E-02&-0.267E-02& 0.104E-01&-0.223E-02 \\
            & 300000.& 0.106E-01&-0.298E-02& 0.108E-01&-0.249E-02 \\
            & 500000.& 0.113E-01&-0.338E-02& 0.112E-01&-0.280E-02 \\
\hline

  O III     &  20000.& 0.747E-02&-0.366E-02& 0.982E-02&-0.340E-02 \\
  4082.2 \AA  &  50000.& 0.133E-01&-0.656E-02& 0.150E-01&-0.583E-02 \\
 C= 0.22E+21& 100000.& 0.166E-01&-0.886E-02& 0.173E-01&-0.733E-02 \\
 3p$'^3$D-3s$'^3$P$^{\degr}$& 200000.& 0.195E-01&-0.106E-01& 0.196E-01&-0.881E-02 \\
            & 300000.& 0.213E-01&-0.117E-01& 0.207E-01&-0.980E-02 \\
            & 500000.& 0.234E-01&-0.133E-01& 0.217E-01&-0.108E-01 \\
\hline

  O III     &  20000.& 0.131E-01&-0.336E-02& 0.162E-01&-0.313E-02 \\
  4110.8 \AA  &  50000.& 0.208E-01&-0.610E-02& 0.226E-01&-0.543E-02 \\
 C= 0.38E+21& 100000.& 0.243E-01&-0.832E-02& 0.257E-01&-0.688E-02 \\
 3d$'^5$P-3p$'^5$S$^{\degr}$& 200000.& 0.275E-01&-0.997E-02& 0.284E-01&-0.831E-02 \\
            & 300000.& 0.291E-01&-0.110E-01& 0.293E-01&-0.914E-02 \\
            & 500000.& 0.307E-01&-0.124E-01& 0.304E-01&-0.102E-01 \lstrut \\
\hline
\end{tabular}
}}
\end{center}
\vskip-1mm
\end{table}

In Djeni\v {z}e et al. (2002) and Milosavljevi\'{c} et al. (2001) we determined Stark broadening parameters
due to collisions with electrons for 15 Ne II and 3 Ne III, and in Sre\'{c}kovi\'{c} et al. (2001) electron-impact widths
for 5 O III multiplets. However, for stellar atmospheres research, Stark broadening data due to
collisions with the principal ionic perturbers, protons and ionized helium, are also useful.
In order
to complete data to be included in STARK-B database, we determined here these additional data.
Also,  we determined within the semiclassical perturbation method electron, proton, and ionized
helium impact broadening parameters for the important Ne II  2s$^2$2p$^4(^{3}$P)3p$^2$D$^{\degr}$ - 2s$^2$2p$^4(^{3}$P)3d$^2$P
multiplet in the visible part of the spectrum, and the missing data for 3 O III multiplets.

\sectionb{2}{RESULTS AND DISCUSSION}

Stark broadening parameters have been determined within the semiclassical perturbation formalism, discussed in detail in Sahal-Br\'echot (1969ab). The optimization and updates can be found in e.g. Sahal-Br\'echot (1974), Dimitrijevi\'{c} \& Sahal-Br\'echot (1984). All details of the calculation are in Djeni\v {z}e et al. (2002) for Ne II and Ne III and in Sre\'{c}kovi\'{c} et al. (2001) for O III, and will not be repeated here.
Obtained results for electron, proton, and ionized helium impact broadening parameters for Ne II, Ne III are presented in Tables 1-3. Table 4 contains the corresponding results for O III. Results are obtained for a perturber density of 10$^{17}$ cm$^{-3}$.

\sectionb{3}{CONCLUSION}
 Using the semiclassical perturbation theory and the corresponding computer code,  we determined Stark broadening parameters due to collisions with protons and helium ions for 15 Ne II and 5 O III multiplets, needed for stellar plasma research, missing in Djeni\v {z}e et al. (2002), Milosavljevi\'{c} et al. (2001) and Sre\'{c}kovi\'{c} et al. (2001), where only electron-impact broadening data are given. The missing data are also needed for the STARK-B database (http://stark-b.obspm.fr), containing Stark broadening parameters for modeling stellar atmospheres, for stellar spectra analysis and synthesis, for the laboratory plasma, inertial fusion plasma, laser development and plasmas in technology investigations. It is a part of Virtual Atomic and Molecular Data Center - VAMDC, an European FP7 project  (P.I. Marie Lise Dubernet, Dubernetet al., 2010, Rixon et al., 2010),  with aims: (i) to build a secure, flexible and interoperable e-science environment based interface to the existing Atomic and Molecular  databases; (ii) to coordinate groups involved in the generation, evaluation, and use of atomic and molecular data, and (iii) to  provide a forum for training of potential users.

Also,  we determined  electron, proton, and ionized
helium impact broadening parameters for the important Ne II  2s$^2$2p$^4$($^3$P)3p$^2$D$^{\degr}$ - 2s$^2$2p$^4$($^3$P)3d$^2P$

multiplet in the visible part of the spectrum, and for 3 O III multiplets.

Due to the high abundance of neon and oxygen in stellar atmospheres, we hope that the obtained results will be of interest for interpretation and synthesis of stellar spectra and for a number of other investigations in astrophysics, physics and plasma technologies.

\newpage
\thanks{This work has been supported by VAMDC,  funded under the "Combination of Collaborative Projects and Coordination and  Support Actions" Funding Scheme of The Seventh Framework Program. Call topic: INFRA-2008-1.2.2 Scientific Data Infrastructure. Grant Agreement number: 239108.  The authors are also  grateful for the support  provided by Ministry Education and Science of Republic of Serbia through project  176002 "Influence of collision processes on astrophysical plasma spectra".}

\References

\refb Djeni\v ze S., Milosavljevi\'{c} V., Dimitrijevi\'{c} M. S. 2002, A\&A, 382, 359

\refb Dimitrijevi\'{c} M. S., Sahal-Br\'echot S. 1984, JQSRT, 31, 301

\refb Dubernet M. L. et al. 2010, JQSRT, 111, 2151

\refb Milosavljevi\'{c} V., Dimitrijevi\'{c} M. S., Djeni\v ze S. 2001, ApJS, 135, 115

\refb  Rixon G., Dubernet M. L., Piskunov N. et al., 2011, AIP Conference Proceedings 1344, 107

\refb Sahal-Br\'echot S. 1969a, A\&A, 1, 91

\refb Sahal-Br\'echot S. 1969b, A\&A, 2, 322

\refb Sahal-Br\'echot S. 1974, A\&A, 35, 321

\refb Sre\'{c}kovi\'{c} V., Dimitrijevi\'{c} M. S.,  Djeni\v {z}e S. 2001, A\&A, 371, 354

\end{document}